\documentclass[namedreferences]{solarphysics}

\usepackage[hyperref,optionalrh]{spr-sola-addons} % For Solar Physics 
\usepackage{graphicx}        % For eps figures, newer & more powerful
\usepackage{color}           % For color text: \color command
\usepackage{breakurl}        % For breaking URLs easily trough lines
\newcommand{\hatssm}{HATS-smm}
\newcommand{\hatsir}{HATS-mir}
\newcommand{\hics}{\textsf{HICS}}
\newcommand{\homs}{\textsf{HOMS}}

\newcommand{\Fqs}{F_\mathrm{QS}}

\newcommand{\Ffl}{F_\mathrm{FL}}
\newcommand{\Tsk}{T_\mathrm{SK}}
\newcommand{\Fsk}{F_\mathrm{SK}}

\newcommand{\Fob}{F_\mathrm{obs}}
\newcommand{\Fmin}{F_\mathrm{min}}
\newcommand{\Pmin}{P_\mathrm{min}}
\newcommand{\Pnoise}{P_\mathrm{noise}}
\newcommand{\kb}{k_\mathrm{B}}
\newcommand{\cf}{\nu_\circ}
\newcommand{\cl}{\lambda_\circ}
\newcommand{\um}{\mu\mathrm{m}}
\newcommand{\uW}{\mu\mathrm{W}}

\newcommand{\aap}{    {\it Astron. Astrophys.}}

\newcommand{\aapr}{   {\it Astron. Astrophys. Rev.}}

\newcommand{\appop}{\textit{App. Optics}}
\newcommand{\apj}{    {\it Astrophys. J.}}
\newcommand{\apjl}{   {\it Astrophys. J. Lett.}}

\newcommand{\araa}{   {\it Ann. Rev. Astron. Astr.}}

\newcommand{\jastp}{  {\it J. Atmos. Sol-Terr. Phys.}}

\newcommand{\solphys}{{\it Solar Phys.}}

\chardef\us=`\_

\begin{document}
\begin{article}
\begin{opening}

\title{HATS: A Ground-Based Telescope to Explore the THz Domain}

\author[addressref={aff1,aff2},corref,email={guigue@craam.mackenzie.br}]{\inits{C.G.}\fnm{C. Guillermo}~\lnm{Gim\'enez
    de Castro}\orcid{0000-0002-8979-3582}}
\author[addressref={aff1}]{\inits{J.-P.}\fnm{Jean-Pierre}~\lnm{Raulin}\orcid{0000-0002-7501-3231}}
\author[addressref={aff1}]{\inits{A.}\fnm{Adriana}~\lnm{Valio}\orcid{0000-0002-1671-8370}}
\author[addressref={aff1}]{\inits{G.}\fnm{Guilherme}~\lnm{Alaia}}
\author[addressref={aff3}]{\inits{V.}\fnm{Vinicius}~\lnm{Alvarenga}}
\author[addressref={aff4}]{\inits{E.C.}\fnm{Emilio~Carlos}~\lnm{Bortolucci}\orcid{0000-0001-8932-2169}}
\author[addressref={aff1}]{\inits{S.H.}\fnm{Silvia~Helena}~\lnm{Fernandes}\orcid{0000-0003-3138-8872}}
\author[addressref={aff5}]{\inits{C.}\fnm{Carlos}~\lnm{Francile}}
\author[addressref={aff1}]{\inits{T.}\fnm{Tiago}~\lnm{Giorgetti}}
\author[addressref={aff1}]{\inits{A.S.}\fnm{Amauri~Shossei}~\lnm{Kudaka}\orcid{0000-0002-5950-7135}}
\author[addressref={aff1}]{\inits{M.F.}\fnm{Fernando~Marcelo}~\lnm{L{\' o}pez}\orcid{0000-0002-2047-6327}}
\author[addressref={aff6,aff7}]{\inits{R.}\fnm{Rog\'erio}~\lnm{Marcon}}
\author[addressref={aff8}]{\inits{A.}\fnm{Adolfo}~\lnm{Marun}\orcid{0000-0002-5738-5648}}
\author[addressref={aff9}]{\inits{M.}\fnm{M\'arcio}~\lnm{Zaquela}\orcid{0000-0002-0419-3204}}

\address[id=aff1]{Centro de R\'adio Astronomia e Astrof\'isica
  Mackenzie, Escola de Engenharia, Universidade Presbiteriana
  Mackenzie, Rua da Consolac\~ao 896, 01302-907, S\~ao Paulo, Brazil.}

\address[id=aff2]{Instituto e Astronom\'ia y F\'isica del Espacio,
  CONICET-UBA, CC. 67 Suc. 28, 1428, Buenos Aires, Argentina.}

\address[id=aff3]{Mackgraphe, Escola de Engenharia, Universidade Presbiteriana
  Mackenzie, Rua da Consolac\~ao 896, 01302-907, S\~ao Paulo, Brazil.}

\address[id=aff4]{Centro de Componentes Semicondutores e Nanotecnologias, Unicamp, Campinas, Brazil.}

\address[id=aff5]{Observatorio Astron\'omico F\'elix Aguilar, Universidad Nacional de San Juan, San Juan, Argentina.}

\address[id=aff6]{Instituto de F\'{\i}sica Gleb Wataghin, Universidade
  Estadual de Campinas, Campinas, SP, Brazil.}

\address[id=aff7]{Observat\'orio Solar Bernard Lyot, Campinas,
  SP, Brazil.}

\address[id=aff8]{Instituto de Ciencias Astron\'omicas, de la Tierra y
  del Espacio, CONICET-UNSJ, San Juan, Argentina.}

\address[id=aff9]{Advanced Embedded R\&D Lab, S\~ao Paulo, Brazil.}

\runningauthor{C.G. Gim\'enez de Castro \textit{et al.}}
\runningtitle{HATS: A Ground-Based Telescope to Explore the THz Domain}

\begin{abstract}
  The almost unexplored frequency window from submillimeter to
  mid-infrared (mid-IR) may bring new clues about the particle
  acceleration and transport processes and the atmospheric thermal
  response during solar flares.  Because of its technical complexity
  and the special atmospheric environment needed, observations at
  these frequencies are very sparse.  The High Altitude THz Solar
  Photometer (HATS) is a full-Sun ground-based telescope designed to 
  observe the continuum from the submillimeter to the mid-IR. It
  has a 457-mm spherical mirror with the sensor in its primary
  focus. The sensor is a Golay cell with high sensitivity in a very
  wide frequency range. The telescope has a polar mount, and a
  custom-built data acquisition system based on a 32 ksamples per
  second, 24 bits (72 dB dynamic range), 8 channels analog-to-digital
  board. Changing only the composition of the low- and band-pass
  filters in front of the Golay cell, the telescope can be setup to
  detect very different frequency bands; making the instrument very
  versatile. In this article we describe the telescope characteristics
  and its development status. Moreover, we give estimates of the
  expected fluxes during flares.
  
\end{abstract}
\end{opening}

\section{Introduction}

Until the installation of the Solar Submillimeter Telescope
\citep[SST:][]{Kaufmannetal:2008} in 1999, flare radiation in the
submillimeter wavelength range was almost unexplored. SST has shown
that the gyrosynchrotron emission model applied for microwaves
\citep{Bastianetal:1998} does not always fit in the submillimeter
wavelength range observations. Indeed,
\cite{Kaufmannetal:2004} have shown for the first time a flare with a
submillimeter spectrum inverting the tendency to decrease with
frequency as expected for the optically thin gyrosynchrotron emission
\citep{Ramaty:1969}. More observations
\citep{Luthietal:2004a,Silvaetal:2007,Kaufmannetal:2009} confirmed the
existence of these new \textit{THz-events}. A review of the observed
flares at the submillimeter range can be found in
\cite{Kruckeretal:2013} and \cite{Fernandesetal:2017}. It is clear,
that there is not one unique mechanism that can explain the emission
at this frequency range, and that observations at higher frequencies
are needed to complete the observational diagnostics.\\

However, the nearer the THz limit, the higher the atmospheric opacity
and the costs associated to build and operate such
facility. At the other end of the THz range, the infrared
(IR), commercial cameras at the focus of small telescopes can be used
to image flares with a relatively good
sensitivity. \cite{Kaufmannetal:2013} have shown at the mid-IR
frequency of 30\,THz ($10\ \um$) strong flux densities during an
X-class flare which are temporally coincident with the white-light
flare.  The origin of this emission was interpreted by
\cite{Trottetetal:2015} as the atmospheric thermal response to the
energy deposited by impinging particles (electrons and ions) over the
chromosphere. Later on \cite{Pennetal:2016} showed that weaker flares,
in the Geostationary Operational Environmental Satellite (GOES) class C7,
also produce mid-IR radiation.\\

%%%%%%%%%%%%%%%%%%%%%%%%%%%%%%%%%%%%%%%%%%%%%%%%%%%%%%%%%%%%%%%%

The IR emission mechanism during flares was first described by
\cite{OhkiHudson:1975} who related it with the apperance of
whithe-light flares and suggested two different sources: a
chromospheric optically thin thermal bremsstrahlung and/or a
photospheric optically thick black body emission dominated by the
H$^-$ opacity. \cite{Kasparovaetal:2009} and \cite{Simoesetal:2017}
arrived to similar conclusions modelling the IR to mm-waves emission
starting from a non-perturbed atmosphere and calculating the
ionization produced by the precipitating electrons via hydrodynamics
simulations; while \cite{HeinzelAvrett:2012} reached similar results
using \cite{Machadoetal:1980} and \cite{Mauasetal:1990} semi-empirical
flare models.  Moreover, \cite{Kasparovaetal:2009} showed that the
thermal radiation is modulated by the beam flux and can be composed of
very fast pulses. \cite{Simoesetal:2017} found that the near IR
continuum is optically thin, as shown by \cite{Pennetal:2016} with
observations at 5.2 and 8.2~$\mu$m and by \cite{Trottetetal:2015} at
10~$\mu$m. Moreover, Sim\~oes showed that for $\lambda\simeq 50\ \mu$m
the emission becomes optically thick.  These results are, however,
model-dependent, therefore a better observational spectral coverage
between the millimeter range and IR is needed to properly assess the
theoretical models, characterize the radiation mechanism, and reveal
the process of energy transport from the energy release site to the
radiation emission location. Since the observation of powerful flares
in Proxima Centauri at 0.233\,THz
\citep{MacGregoretal:2018,MacGregoretal:2020} the interest in this
spectral range is not no longer just relevant to the Sun.\\

%%%%%%%%%%%%%%%%%%%%%%%%%%%%%%%%%%%%%%%%%%%%%%%%%%%%%%%%%%%%%%%%%%%%%%%%%%%%

The Solar-T balloon experiment \citep{Kaufmannetal:2014} was the first
instrument designed to observe the Sun in the THz range. This article
describes new ground-based telescopes for observations at the
submillimeter and mid-IR wavelengths similar to Solar-T. In this
article we explain the particularities of these new facilities, the
expected fluxes, and put them in the context of previous works.\\

\section{HATS Description}
\label{sec:hats}

The telescope concept is described by \cite{Kaufmannetal:2014}; it is
based on a single opto-acoustic photometer (a Golay cell) at the focus
of a reflector telescope. The optical system creates an image of the
Sun with the size of the entrance cone of the Golay cell, i.e. even
with a large mirror and only one-pixel sensor, the system behaves like
a full-Sun instrument. This concept was applied in the Solar-T balloon
experiment that flew over Antarctica for two weeks in January 2016
with a couple of Cassegrain telescopes and receivers for 3 and 7 THz
respectively.\\

\begin{figure} %%%%%%%%%%%%%%%%%% FIGURE 1
  \centerline{\includegraphics[width=0.7\textwidth]{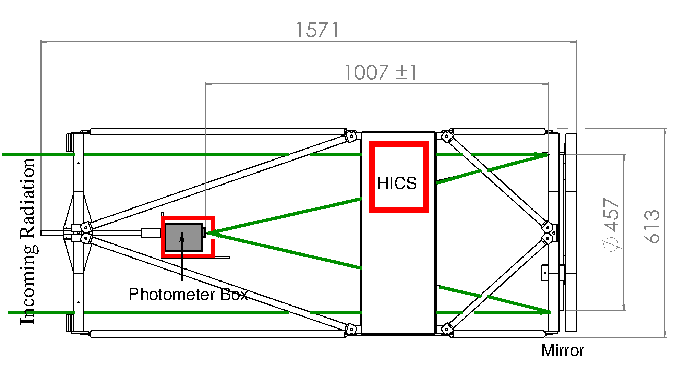}\hspace{0.7cm}
    \includegraphics[width=0.4\textwidth,angle=90,origin=lb]{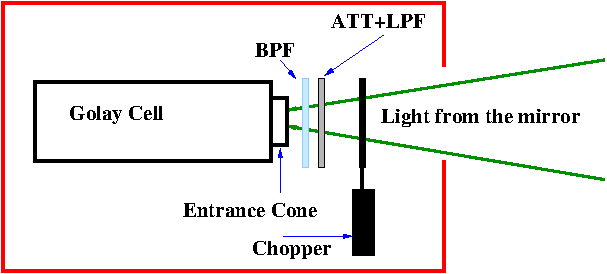}}
  \caption{\textit{Left}: HATS general diagram, a prime focus system
    composed of a 457\,mm spherical mirror, the photometer box, and
    the data acquisition computer HATS Interface and Control System
    (\hics). Dimensions are in mm. \textit{Right}: The Photometer box
    with the chopper, attenuators (ATT), low-pass (LPF) and band-pass
    (BPF) filters. Green lines represent light rays.}
  \label{fig:HATS-diagram}
\end{figure}

HATS is a ground-based telescope that uses the same concept
but is designed to operate from the submillimeter to the mid-IR. It is
a prime focus system, with a spherical mirror.  To reduce the IR
radiation the mirror is roughened. It has a diameter $\phi = 457$\,mm
and a focal distance $f = 1007$\,mm (Figure \ref{fig:HATS-diagram}),
therefore the optical solar disk will produce an image at the focus
with diameters between $9.21$\,mm (aphelion), $9.52$\,mm (perihelion)
and $9.36$\,mm in average, all of them smaller than the photometer
entrance cone which has a diameter $\ge 10$\,mm.  The mirror was
roughened using carborundum $1.25\ \mu\mathrm{m}$ (E10); its
reflectance $R(\lambda)$ in function of wavelength yields $R\simeq 1$
and $R\simeq 0.3$ for $\lambda=300\ \mu\mathrm{m}$ and
$\lambda=20\ \mu\mathrm{m}$, respectively
\citep{Fernandes.Master:2013}. The Golay cell is inside a box, with
the chopper, low-pass filters, attenuators, and the frequency selector
band-pass filter. The set of filters can be changed to select
different wavelength ranges. Attenuators and low-pass filters are used
to block $\lambda < 5 \ \mu\mathrm{m}$ wavelengths and reduce the
incoming power to the sensor maximum measurable power. Figure
\ref{fig:HATS-diagram} shows diagrams of the telescope and the
photometer box where the Golay cell, filters, and chopper are
installed.\\

HATS was originally conceived as a robotic telescope observing at
$\cf=0.87$\,THz and $\cf=1.4$\,THz or $\cl=344\ \um$ and
$\cl=214\ \um$, respectively \citep{Kaufmannetal:2015b}. It has two
metal-mesh band-pass filters placed in a rotating wheel with period
$P\simeq 5$~Hz that also serves as a chopper. They were built at
Centro de Componentes Semicondutores (CCS), a laboratory of the
Universidade Estadual de Campinas (Unicamp) and tested at the Max
Planck Institute f\"ur extraterrestrische Physik in Garching, Germany
\citep{Meloetal:2008}.  Filter characteristics are presented in Table
\ref{tbl:BPF}. In order to minimize restrictions from atmospheric
absorption, the telescope should be installed at a very high altitude
($\ge 5000$\,meters above sea level (masl)) where the precipitable
water vapor content (PWV) should be less than 1\,mm during a large
fraction of the year. Since such locations are generally isolated with
limited infrastructure, like Solar-T, HATS should use satellite
communication for data downloading and monitoring, electric power from
solar panels and an intelligent radome that opens during observations
and closes at night or when atmospheric conditions may damage the
system.\\

While we look for a place where to install the THz telescope, we are
working in a mid-IR version of HATS because the atmospheric opacity at
this frequency is not very high at medium elevations, and therefore
there are more candidates for the installation site.  To distinguish
between the different HATS setups we refer to the submillimeter
version as the \hatssm\ and the mid-IR as \hatsir.\\

\begin{figure} %%%%%%%%%%%%%%%%%% FIGURE 2
  \centerline{\includegraphics[width=0.45\textwidth]{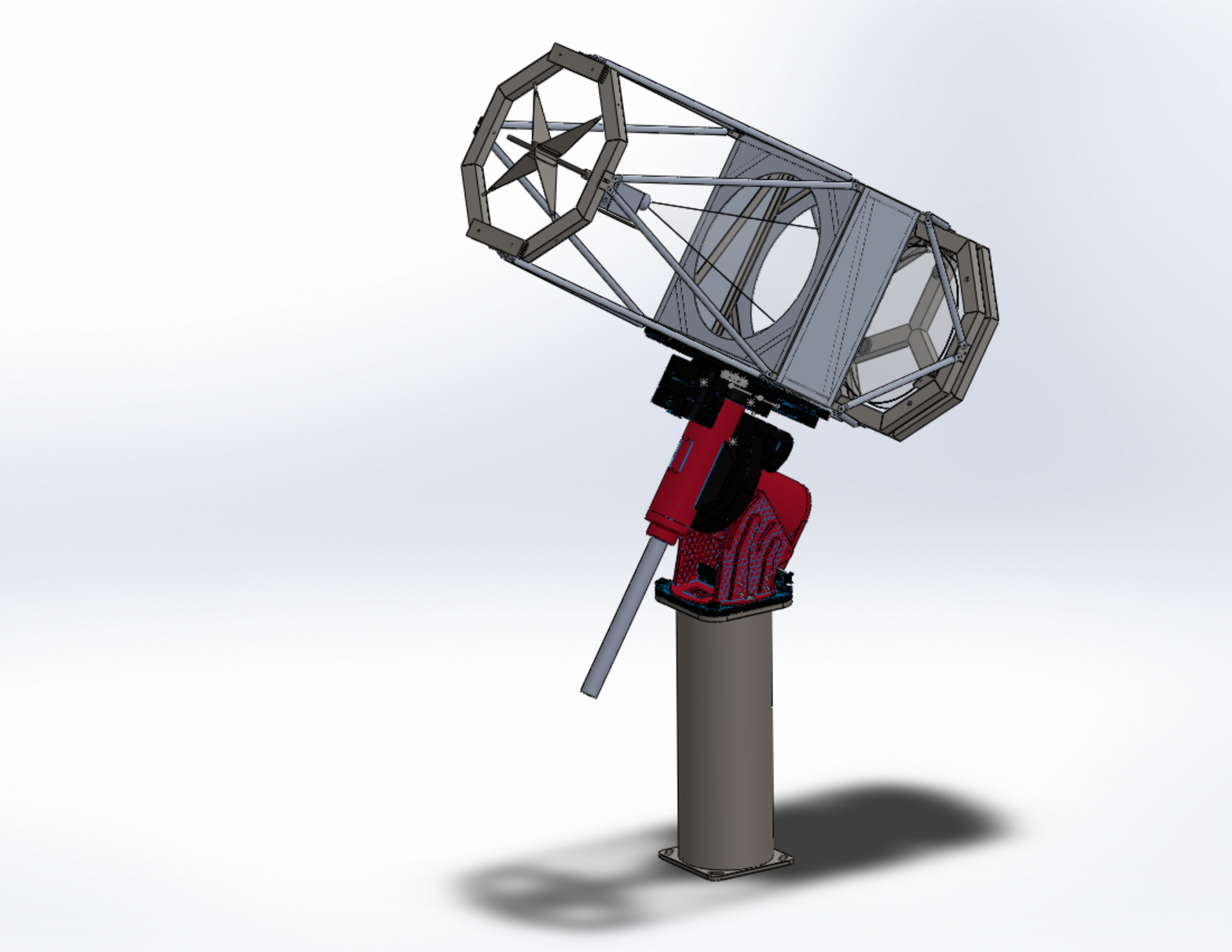}\hfill
  \includegraphics[width=0.45\textwidth]{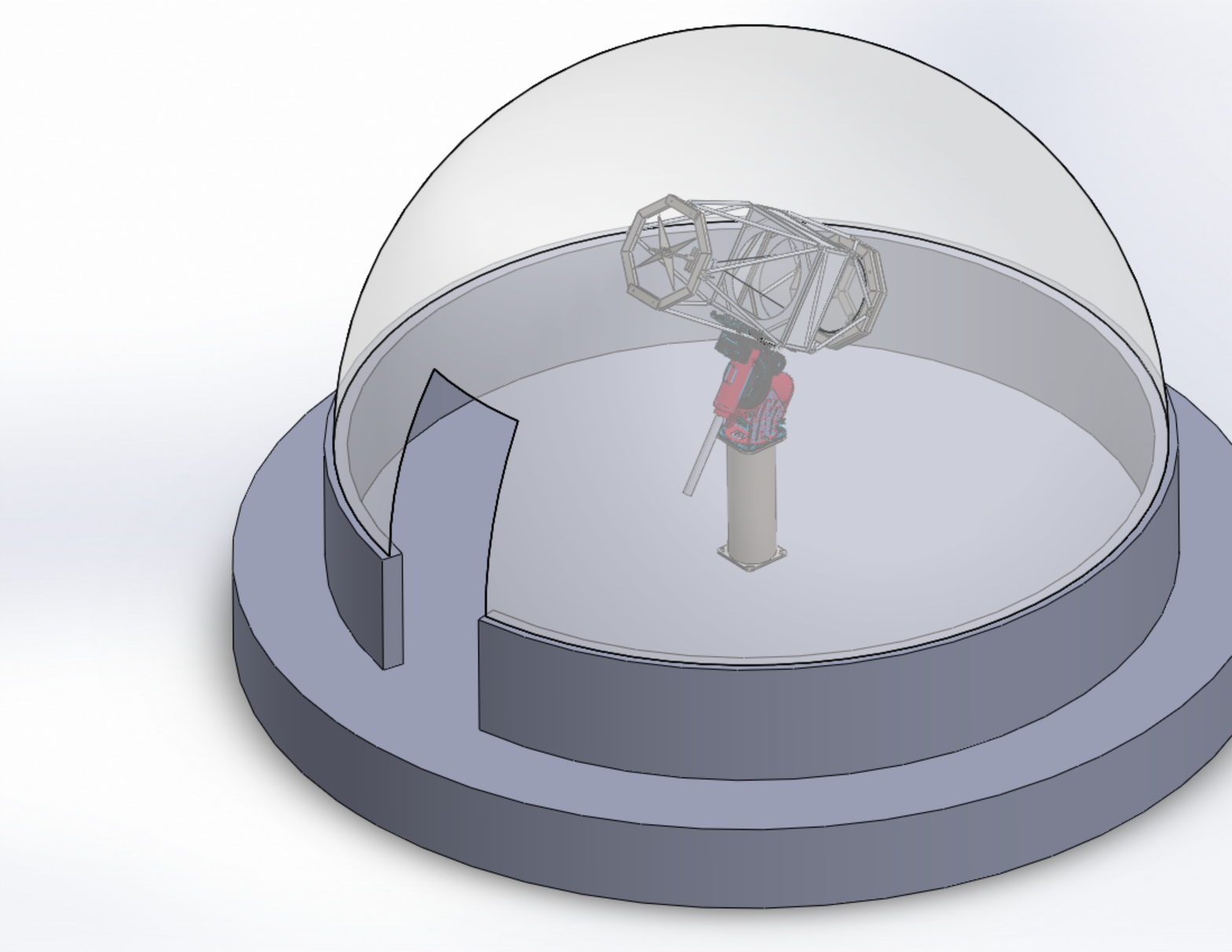}}
\caption{A projected 3D view of \hatsir\ (left) and inside the
polypropylene radome in park position, pointing to South (right).}
  \label{fig:HATS-3D}
\end{figure}

\begin{figure} %%%%%%%%%%%%%%%%%% FIGURE 3
  \centerline{\includegraphics[width=0.8\textwidth]{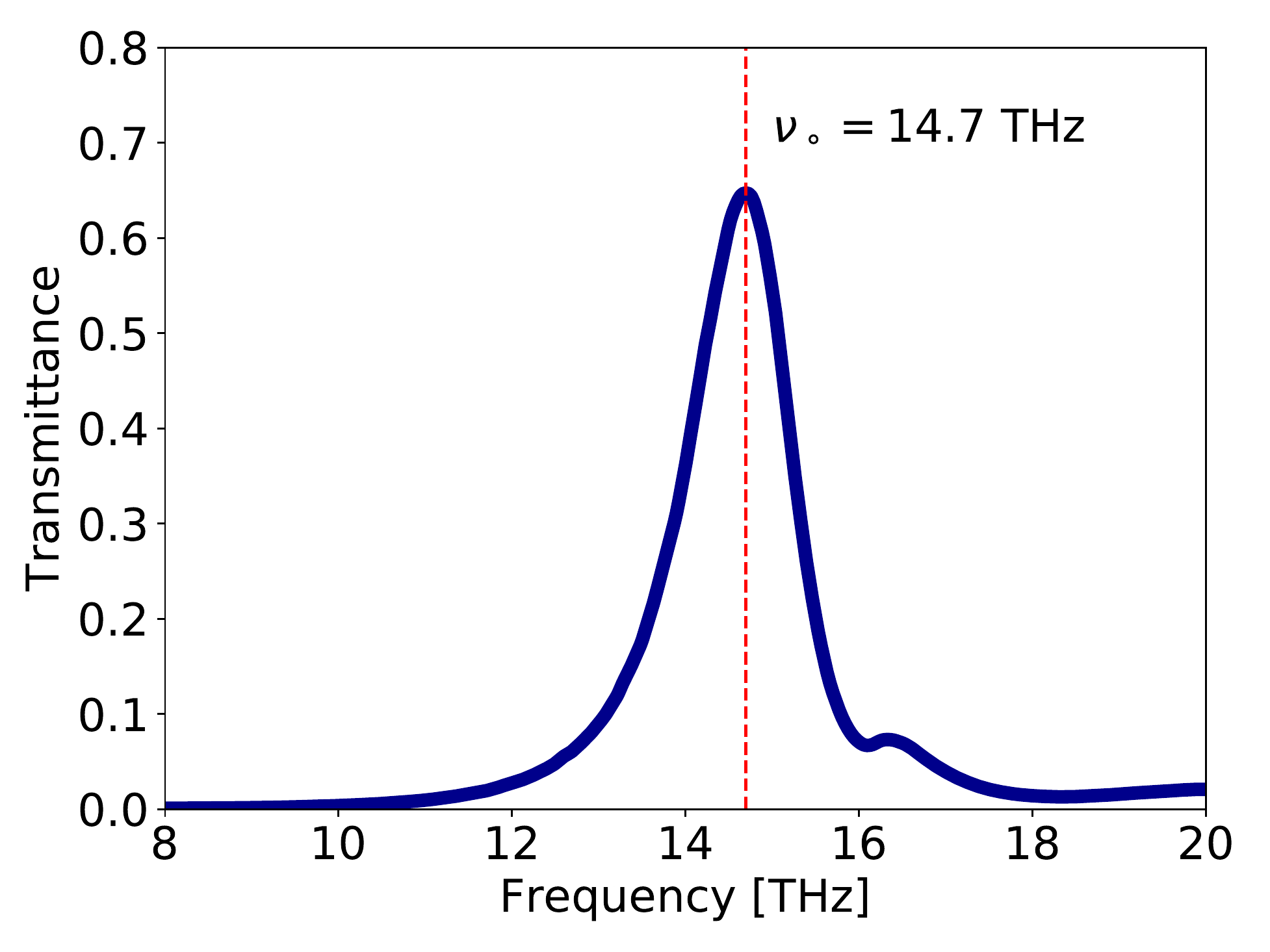}}
  \caption{Transmittance of two stacked Tydex band-pass filters as a 
    function of frequency.}
  \label{fig:Tydex-bpf}
\end{figure}

\hatsir\ has a sensor with a central frequency $\cf=15$\,THz
($\cl=20\,\um$), see Table \ref{tbl:BPF}.  It will be installed at
$\simeq 2300$\,masl at the Observatorio Astron\'omico F\'elix Aguilar
(OAFA, Argentina) which will provide electric power and Internet
connection; the telescope will be enclosed in a radome and manually
operated from a control room.  Figure \ref{fig:HATS-3D} is a projected
3D view of \hatsir\ including the radome. The telescope is made of
aluminum bars, has lateral bays to accommodate auxiliary devices, is
placed over a Paramount ME\,\textsc{ii} equatorial mount and protected
from the wind and dust by a translucent polypropylene semi-spherical
radome 4.5~m wide standing over a 0.7\, m high circular wall. The
polypropylene material was chosen because it has a relatively high
transmission ($\simeq 0.35$) at the observing frequency. The decision
as to whether we will use a fork or a wheel chopper (in Figure
\ref{fig:HATS-diagram} a fork chopper is represented) is still being
evaluated. The band-pass filter is fabricated from thin metal foil
with holes by the Saint Petersburg company Tydex, model BPF15.0-24,
attenuators and low-pass filters are also fabricated by Tydex. We plan
to stack together two band-pass filters to increase the side frequency
rejection, Figure \ref{fig:Tydex-bpf} shows the response in frequency
of the set of filters. The responses of all components in the
telescope light path are tested with a Fourier Transform Infrared
Interferometer in the Centro de Pesquisas em Grafeno e Nanomateriais
(Mackgraphe) of the Universidade Presbiteriana Mackenzie, S\~ao Paulo,
Brazil. Some of the optical system components may change until the
telescope is finally installed, however, the total transmission
$\eta_{\nu_\circ}$ should not change significantly because we can
compensate using different optical attenuators.\\

\begin{table}
    \begin{tabular}{lllll}
      \textbf{Origin} & \textbf{Wavelength} & \textbf{Frequency} & \textbf{Bandwidth}         & \textsf{Transmission} \\
                      &  [$\mu\mathrm{m}$]  &[THz]               &  [\% of central Frequency] & [\%] \\
      \hline
      CCS             &  352                & 0.87               &        14.0                &  84 \\
      \textsf{Tydex}$\dag$  &  20.0         & 15.0               &         7.0                &  65 \\
      \hline
      \multicolumn{3}{l}{$\dag$: two stacked filters.}
    \end{tabular}
    \caption{Characteristics of the band-pass filters.}
    \label{tbl:BPF}
\end{table}

Both \hatsir\ and \hatssm\ use a Tydex Golay cell GC-1T whose analog
output is digitized using an AD7770 Analog Devices Inc. converter with
24 bits, 8 channels and 32\,ksamples\,s$^{-1}$.  The converter is
controlled with a ATMEL ATSAME-XPRO single board computer running a
real time Linux operating system (\textsf{freeRTOS}): this ensemble is
called the HATS Interface and Control System (\textsf{HICS}) and is
installed in the telescope lateral bay (Figure
\ref{fig:HATS-diagram}). With a 24-bit analog-to-digital converter, we
have a 72 dB dynamic range, which allows us to detect signals much
weaker than the sensor sensitivity. The HATS Operation, Monitoring and
Storage (\textsf{HOMS}) is a desktop computer installed at the control
room running the telescope control software, connected through
Ethernet to \hics\ and the Paramount mount. The mount has a
proprietary control software called \textsf{TheSkyX}. While this
software has a graphical user interface, remote commands, written in
Javascript, can be sent to the mount through an Ethernet
connection. Global Positioning System (GPS) receiver is used to set
\homs\ time, that runs an Netwrok Time Protocol (NTP) server to
synchronize the computers in the network.  Data are read at a 1 kHz
rate, and transferred to \textsf{HOMS} where it is window-Fourier
transformed. A peak around the chopper frequency should show up,
therefore, the window of the Fourier transforms defines the signal
time resolution: from 256 to 1024 ms. The sensor response \textbf{as
  a} function of temperature is calibrated by comparing the output
signal against a black-body source, resulting in a linear relationship
between voltage and temperature which is later used for data
calibration \citep{Kaufmannetal:2014}.

\section{Expected Signal Input}

Here we compute the expected incoming power for $\nu=0.87$\, THz and
$15$\, THz.  There are three sources of emission: the quiet Sun (QS),
the flare (FL), and the sky (SK), the observed flux density $\Fob$ will
then be given by
\begin{equation}
 \Fob = \Ffl + \Fqs + \Fsk \ .
\end{equation}
We will analyze each term separately.  First the QS; since the Sun has
a brightness temperature between 4500 and 5500\,K for the observing
range of the telescope, the Raleigh-Jeans (RJ) approximation is valid
for the submillimetric to the mid-IR range.  Therefore the expected
flux density above the atmosphere is
\begin{equation}
  \Fqs = \frac{2 \kb T_\odot \cf^2}{c^2} \Omega_\odot \quad [\mathrm{W m^{-2} Hz^{-1}}] \ ,
\end{equation}
where $\kb$ is the Boltzmann constant, $cf$ the speed of light in
vaccum, $T_\odot$ is the QS temperature for the frequency $\cf$ and
$\Omega_\odot$ is the Sun solid angle from Earth, which is
smaller than the telescope field of view solid angle
$\Omega_\mathrm{t}$. The sky contribution to the flux density is, on
the other hand,
\begin{equation}
  \Fsk = \frac{2 \kb \Tsk \cf^2}{c^2} \Omega_\mathrm{t} \left ( 1 - e^{-\tau_{\nu_\circ}\zeta}\right ) \quad [\mathrm{W m^{-2} Hz^{-1}}] \ ,
\end{equation}
where $\Tsk$ is the sky brightness temperature, $\tau_{\nu_\circ}$ the zenith
optical depth for the frequency $\nu_\circ$, and
$\zeta = 1/\sin(\mathrm{elevation})$, the airmass.\\

Estimating $\Ffl$ is model dependent. For instance, we can use
\cite{Simoesetal:2017} to deduce brightness temperature for a
\textit{thermal} flare.  However we prefer to determine the minimum
flux density that could be detected. This will be addressed in the
next section.\\

$\Fqs$ and $\Ffl$ are attenuated by the atmosphere by a factor
\begin{equation}
  {\cal A}(\nu) = e^{-\tau_{\nu_\circ} \zeta} \ .
\end{equation}
We set in this analysis $\zeta=1.71$ which corresponds to the mean
value for elevations between $10^\circ$ and $85^\circ$. Values for
$\tau_{\nu_\circ}$ are strongly dependent on the observing site. We
consider in this work two different absorption scenarios: \textit{i)}
a mid-altitude site at 2500\, masl for $\cf=15$\,THz and \textit{ii)}
a high-altitude site at 5000\, masl for $\cf=0.87$\,THz. The mid-IR
camera report of the Gran Telescopio Canarias
(\url{http://www.gtc.iac.es/instruments/canaricam/MIR.php}, accessed
on 2019-11-19) uses Atmospheric Transmission (ATRAN) models for
atmospheric transmission in the range $300 \ge \nu \ge 12$\,THz at a
mid-altitude site. Results imply that the transmission is weakly
dependent on PWV, for a range between 2.3 to 10\, mm, yielding a mean
value $\tau_\nu=0.35$ for $12 \le \nu \le 20$\,THz. The PWV range used
for the ATRAN simulations is similar to the range at the Complejo
Astron\'omico El Leoncito (CASLEO) observatory
\citep{Cassianoetal:2018}, which is 5\,km away from OAFA where
\hatsir\ will be installed.\\

On the other hand, the submillimeter wavelengths are strongly
dependent on PWV. Using the Atmospheric Transmission at Microwaves
(\textsf{ATM}) model \citep{Pardoetal:2001} and the statistical
analysis of the optical depth at 210 GHz for the Alto Chorrillo site
\citep[4850\,masl, Salta, Argentina, ][]{Bareillesetal:2011}, we
obtain a $\tau_\nu\le 1.0$ at $\nu=0.87$ THz for around 30 days per
year.\\

To get the total power received by the photometer $P_{\nu_\circ}$, we
need to multiply by the mirror surface area $A_\mathrm{surf}$, the
product of all of the blocking, transmission, and reflection factors
along the optical path $\eta_{\nu_\circ}$ and convolve with the
band-pass filter response $f_\mathrm{PB}(\nu)$, i.e.
\begin{equation}
  P_{\nu_\circ} = \frac{A_\mathrm{surf} \, \eta_{\nu_\circ}}{2} \int \Fob(\nu) f_\mathrm{BP}(\nu) \mathrm{d}\nu \ , \quad
  \Fob = (\Ffl + \Fqs){\cal A}(\nu) + \Fsk \ .
\end{equation}
Table \ref{tbl:setup} summarizes the different parameters used in
these simulations. The total transmission factor $\eta_{\nu_\circ}$
for \hatsir\ is the multiplication of the polypropylene radome
transmission ($0.35$), the mirror reflectivity ($0.3$), the fraction
of mirror light converging to the sensor because of the photometer box
blocking ($\simeq 0.88$), and the chopper ($0.61$ for a fork chopper,
$1.0$ for a wheel chopper). In case the incoming power is still larger
than the Golay cell manufacturer recommended maximum detected power
($10^{-5}$\, W) attenuators must be added in the optical path.\\

\begin{table}
    \begin{tabular}{l|ccccccc}
      \hline\hline
\textbf{Telescope} &      $\nu_\circ$ & $T_\odot$     & $\Tsk$ & $\tau$ & $\zeta$ & $\eta_{\nu_\circ}$ &  $A_\mathrm{surf}$ \\
\textbf{Setup}     & (THz)      &  (K)          & (K)   &         &         &                          &   $(\mathrm{m}^2)$ \\
      \hline
HATS-smm              & 0.87       &  $4500^\dag$  & 290    &  1.0   &  1.71   &    $7.1\times 10^{-1}$& 0.164 \\
HATS-mir              & 15.0       &  $5600^\S$    & 290    &  0.3   &  1.71   &    $1.1\times 10^{-2}$& 0.164 \\
      \hline
    \end{tabular}
    \caption{Table with model parameters. $\dag$:
      \cite{RighiniSimon:1976}, $\S$: \cite{Simoesetal:2017}.}
    \label{tbl:setup}
\end{table}

\subsection{Flare Detectability}

The main purpose of HATS is to characterize flares in the THz to
mid-IR range, and with this aim it was designed.  To estimate the
power produced by a flare in the photometer we will proceed
from the Golay cell noise and convert it to flux density.  The
photometer has a typical noise equivalent power
$\mathrm{NEP}=1.4\times 10^{-4} \,\mu\mathrm{W\ Hz}^{-1/2}$.  The
noise power is determined multiplying the NEP by the square root of the
modulation frequency
\begin{equation}
  \Pnoise = \mathrm{NEP} \sqrt{f_\mathrm{chopper}} =
  1.4\times 10^{-4} \times \sqrt{15} = 5.4\times 10^{-4} \,\uW \ ,
\end{equation}
where $f_\mathrm{chopper}=15$\, Hz is the typical chopper
frequency. By definition $\Pnoise$ is the incoming power on the
detector that produces an output equal to the root mean square (RMS)
noise. To be able to detect an external signal over noise we will
assume a miminum incoming power
$\Pmin = 3 \times \Pnoise = 1.6\times 10^{-3}\,\uW$. Then, the minimum
flare flux density detectable is
\begin{equation}
  \Fmin = \frac{2 \ \Pmin}{A_\mathrm{surf}} \
  \frac{1}{{\cal A}({\nu}) \ \eta_{\nu_\circ} } \
  \frac{1}{\Delta_\mathrm{eff}\nu} \ , \quad
  \Delta_\mathrm{eff}\nu = \int f_\mathrm{BP}(\nu) \ \mathrm{d}\nu \ ,
\end{equation}
$\Delta_\mathrm{eff}\nu$ is the filter-averaged frequency range of the
band-pass filter. Table \ref{tbl:minFx} shows the expected minimum
detectable flux densities for \hatssm\ and \hatsir. We included in the
table the power produced by the quiet Sun and the sky (the total
incoming power should be $\le 10\,\uW$). We note that the dynamic
range needed to detect a flare is
  $$ DR = 10 \ \log_{10}\left ( \frac{P_\mathrm{total}}{\Pmin}\right )\ , \quad
P_\mathrm{Total} = \Pmin+ P_\mathrm{QS} +P_\mathrm{sky} \ .$$
Replacing $P_\mathrm{Total}$ and $\Pmin$ from Table \ref{tbl:minFx} we
get $DR_\mathrm{mir}=38$\,dB and $DR_\mathrm{smm} = 24$\,dB for \hatsir\ and
\hatssm, respectively, which are smaller than the $DR$ 72\,dB of the
acquisition system (Section \ref{sec:hats}).

\begin{table}
  \begin{tabular}{l|cc|ccc|cc}
    \hline\hline
    \textbf{Telescope} & $\cf$  & $\Delta_\mathrm{eff}\nu$ & $\Pmin$  & \textbf{Quiet Sun}& \textbf{Sky} & ${\cal A}(\nu)$ &  $\Fmin$ \\
    \textbf{Setup}     & \multicolumn{2}{c|}{(THz)}         & \multicolumn{3}{c|}{($\uW$)}                 &                 & ($10^3\ $~SFU)  \\
    \hline
    \hatssm            & $0.87$ & $0.11$                   & $0.0016$ & $0.077$           & $0.001$      & $0.18$          &  $14$ \\
    \hatsir            & $15.0$ & $1.19$                   &$0.0016$  & $9.66$            & $0.34$       & $0.60$          &  $23$ \\
    \hline
  \end{tabular}
  \caption{Expected minimum flare flux density detectable for the different HATS setups.}
  \label{tbl:minFx}
\end{table}

\section{Discussion}

The minimum detectable flux densities derived from the telescope
average observing configurations are of the order of
$10-25 \times 10^3$\,SFU. Since there are no previous reports of
flares at the frequencies of HATS operation, we will compare our
estimates with observations at frequencies near those of HATS.\\

In the mid-IR range, \cite{Kaufmannetal:2015} reported flux densities
recorded with a 30\,THz ($10\ \um$) camera as high as
$35\times 10^3$\,SFU for an X2 event (SOL2014-10-27T14:22), the
highest ever reported. Other works also give values of the order of
$10^4$\, SFU, although more moderate
\citep[see][]{Kaufmannetal:2013,GimenezdeCastroetal:2018}. While these
events are strong flares (M8 and X9 GOES class), \cite{Pennetal:2016}
presented observations at $5.2$ and $8.2\ \um$ during the M7 class
flare SOL2014-09-24T17:50 with maximum flux densities $< 1000$\,SFU,
that would not be detected by \hatsir.\\

On the other hand, from theoretical simulations we can reasonably
expect an excess brightness temperature $T_{fl}\simeq 3\times 10^3$\,K
at $\nu\simeq 15$\,THz produced by accelerated electrons heating the
chromosphere \citep{Simoesetal:2017}. Considering an average thermal
source extending over a surface
$A_\mathrm{source} = 6\times 10^{18}\,\mathrm{cm}^2$, corresponding to
a solid angle $\Omega_s \simeq 3\times 10^{-8}$\,str, the flux density
produced by this thermal source would be $F_x = 59\times 10^3$\,SFU,
and therefore, detectable by \hatsir. The main limitation to increase
\hatsir\ sensitivity is the quiet Sun background emission and, to a
lesser extent, the sky radiation: the minimum flare detectable power
is a thousandth of the background power over the photometer. The same
large mirror that allows us to gather enough flux from the tiny flare
to detect it, gathers also the strong mid-IR emission coming from the
full quiet Sun.\\

In the sub-millimeter range, the most intense flare already detected, 
SOL2003-11-02T17:17, an X8 GOES class flare,
had a flux density peak $F_x\simeq 65\times 10^4$\,SFU at $0.4$\,THz
\citep{Silvaetal:2007}. Since its submillimeter spectrum increases
from 0.2 toward 0.4\,THz, it is expected that the flux density at
$0.87$\,THz should be even higher. We note, however, that this was a
rather unusual event with an extremely steep submillimeter spectral
index. Another submillimeter event where the intensity increases
towards the highest frequencies is SOL2013-11-04T19:43
\citep{Kaufmannetal:2004}. During peak time, the submillimeter
spectral index
$$\alpha=\frac{ \log\left ( \frac{F_{0.4}}{F_{0.21}}\right )} {\log\left(\frac{0.4}{0.21}\right )}  \simeq 0.8 \ . $$
If the same spectral index is applied to the flux between $0.4$
and $0.87$\,THz we obtain
$$F_{0.87} = F_{0.4} \left(\frac{0.87}{0.4}\right )^{0.8} \simeq 36\times 10^3 \,\mathrm{SFU} \ . $$
While these numbers may be encouraging, we have to remember that just
a few submillimeter flares had fluxes $> 10^4$\,SFU in the last two
solar cycles \citep{Kruckeretal:2013,Fernandesetal:2017}. The high
atmospheric opacity is the main obstacle at $0.87$\,THz -- if we
consider $\tau_\nu=0.5$ and $\zeta=1.01$ (elevation=$80^\circ$) the
minimum detectable flux density drops to $\simeq 4000$\,SFU. However,
such a low atmospheric optical depth is expected for only a few days
per year in most of the installation sites already analyzed. \\

\hatsir\ is under construction at Centro de R\'adio Astronomia e
Astrof\'{\i}sica Mackenzie (CRAAM, S\~ao Paulo, Brazil) and will be
transported and installed in the OAFA observatory in June 2020. First
light is expected for September 2020. There is still no provision for
the construction of \hatssm.  Both HATS setups, when operating, will
bring flare data from yet unexplored frequencies. These new frequency
windows will complement the millimeter observations of the
POlarization Emission of Millimeter Activity at the Sun \citep[POEMAS,
for 0.045 and 0.090 THz, ][]{Valioetal:2013}, the submillimeter range
of the SST, and the mid-IR of the 30\,THz cameras. As solar dedicated
telescopes, they will create a data base of flares to constrain
theoretical models for the flare energy transport to the
chromosphere/photosphere. We remark that these are unique instruments
in this frequency range, since only the Atacama Large Millimeter Array
(ALMA) can observe solar flares at submillimeter wavelenghts, however
with very limited observing times because of its high demand.\\

%%%%%%%%%%%%%%%%%%%%%%%%%%%%%%%%%%%%%%%%%%%%%%%%%%%%%%%%%%%%%%%%%%%%%%%%%%%
\begin{acknowledgments}

  C.G.~Gim\'enez de Castro and J.-P.~Raulin acknowledge CNPq (contracts
  305203/2016-9 and 312066/2016-3).  The research leading to these
  results has received funding from CAPES grant 88881.310386/2018-01,
  FAPESP grant 2013/24155-3, and the US Air Force Office for Scientific
  Research (AFOSR) grant FA9550-16-1-0072.  

\end{acknowledgments}

\section*{Disclosure of Potential Conflicts of Interest}
The authors declare that they have no conflicts of interest.

\bibliographystyle{spr-mp-sola}
%\bibliography{referencias}  

\begin{thebibliography}{31}
% BibTex style file: spr-mp-sola.bst (nameyear), 2015-03-09
\ifx\bisbn     \undefined \def\bisbn  #1{ISBN #1}\fi
\ifx\binits    \undefined \def\binits#1{#1}\fi
\ifx\bauthor   \undefined \def\bauthor#1{#1}\fi
\ifx\batitle   \undefined \def\batitle#1{#1}\fi
\ifx\bjtitle   \undefined \def\bjtitle#1{\textit{#1}}\fi
\ifx\bvolume   \undefined \def\bvolume#1{\textbf{#1}}\fi
\ifx\byear     \undefined \def\byear#1{#1}\fi
\ifx\bissue    \undefined \def\bissue#1{#1}\fi
\ifx\bfpage    \undefined \def\bfpage#1{#1}\fi
\ifx\blpage    \undefined \def\blpage #1{#1}\fi
\ifx\burl      \undefined \def\burl#1{\textsf{#1}}\fi
\ifx\href      \undefined \def\href#1#2{\textsf{#2}}\fi
\ifx\betal     \undefined \def\betal{\textit{et al.}}\fi
\ifx\bctitle   \undefined \def\bctitle#1{#1}\fi
\ifx\beditor   \undefined \def\beditor#1{#1}\fi
\ifx\bbtitle   \undefined \def\bbtitle#1{\textit{#1}}\fi
\ifx\bedition  \undefined \def\bedition#1{#1}\fi
\ifx\bseriesno \undefined \def\bseriesno#1{\textbf{#1}}\fi
\ifx\blocation \undefined \def\blocation#1{#1}\fi
\ifx\bsertitle \undefined \def\bsertitle#1{\textit{#1}}\fi
\ifx\bsnm      \undefined \def\bsnm#1{#1}\fi
\ifx\bsuffix   \undefined \def\bsuffix#1{#1}\fi
\ifx\bparticle \undefined \def\bparticle#1{#1}\fi
\ifx\barticle  \undefined \def\barticle#1{}\fi
\ifx\binstitute  \undefined \def\binstitute#1{#1}\fi
\ifx\bpublisher  \undefined \def\bpublisher#1{#1}\fi
\ifx\doiurl    \undefined
  \def\doiurl#1{\href{http://dx.doi.org/#1}{\textsf{DOI}}}\fi
\ifx\arxivurl  \undefined
  \def\arxivurl#1{\href{http://arxiv.org/abs/#1}{\textsf{arXiv}}}\fi
\ifx\adsurl    \undefined
  \def\adsurl#1{\href{http://adsabs.harvard.edu/abs/#1}{\textsf{ADS}}}\fi
\ifx\botherref \undefined \def\botherref#1{}\fi
\ifx\url       \undefined \def\url#1{\textsf{#1}}\fi
\ifx\bchapter  \undefined \def\bchapter#1{}\fi
\ifx\bbook     \undefined \def\bbook#1{}\fi
\ifx\bcomment  \undefined \def\bcomment#1{#1}\fi
\ifx\oauthor   \undefined \def\oauthor#1{#1}\fi
\ifx\citeauthoryear \undefined\def \citeauthoryear#1{#1}\fi
\ifx\endbibitem\undefined \def\endbibitem{}\fi
\ifx\bconflocation  \undefined \def\bconflocation#1{#1} \fi

\bibitem[\protect\citeauthoryear{{Bareilles}
  \textit{et~al.}}{2011}]{Bareillesetal:2011}
\begin{bchapter}
\bauthor{\bsnm{{Bareilles}}, \binits{F.}},
\bauthor{\bsnm{{Morras}}, \binits{R.}},
\bauthor{\bsnm{{Hauscarriaga }}, \binits{F.P.}},
\bauthor{\bsnm{{Guarrera}}, \binits{L.}},
\bauthor{\bsnm{{Arnal}}, \binits{E.M.}},
\bauthor{\bsnm{{Lepine}}, \binits{J.R.D.}}:
\byear{2011},
\bctitle{{Comparaci{\' o}n entre los sitios de LLAMA y APEX}}.
In: \bbtitle{Bol. Asoc. Arg. de Astron. 54}.
\end{bchapter}
\endbibitem

\bibitem[\protect\citeauthoryear{{Bastian}, {Benz}, and
  {Gary}}{1998}]{Bastianetal:1998}
\begin{barticle}
\bauthor{\bsnm{{Bastian}}, \binits{T.S.}},
\bauthor{\bsnm{{Benz}}, \binits{A.O.}},
\bauthor{\bsnm{{Gary}}, \binits{D.E.}}:
\byear{1998},
\batitle{{Radio Emission from Solar Flares}}.
\bjtitle{\araa}
\bvolume{36},
\bfpage{131}.
\doiurl{10.1146/annurev.astro.36.1.131}.
\adsurl{https://ui.adsabs.harvard.edu/abs/1998ARA&A..36..131B}.
\end{barticle}
\endbibitem

\bibitem[\protect\citeauthoryear{{Cassiano}
  \textit{et~al.}}{2018}]{Cassianoetal:2018}
\begin{barticle}
\bauthor{\bsnm{{Cassiano}}, \binits{M.M.}},
\bauthor{\bsnm{{Cornejo Espinoza}}, \binits{D.}},
\bauthor{\bsnm{{Raulin}}, \binits{J.-P.}},
\bauthor{\bsnm{{Gim{\'e}nez de Castro}}, \binits{C.G.}}:
\byear{2018},
\batitle{{Precipitable water vapor and 212 GHz atmospheric optical depth
  correlation at El Leoncito site}}.
\bjtitle{\jastp}
\bvolume{168},
\bfpage{32}.
\doiurl{10.1016/j.jastp.2018.01.015}.
\adsurl{https://ui.adsabs.harvard.edu/abs/2018JASTP.168...32C}.
\end{barticle}
\endbibitem

\bibitem[\protect\citeauthoryear{{Fernandes}}{2013}]{Fernandes.Master:2013}
\begin{botherref}
\oauthor{\bsnm{{Fernandes}}, \binits{L.O.T.}}:
2013,
Desenvolvimento de fot{\^o}metros thz para observa{\c c}{\~a}o de explos{\~o}es
  solares.
Master's thesis,
Unicamp.
Link: \url{http://repositorio.unicamp.br/jspui/handle/REPOSIP/259238}.
\end{botherref}
\endbibitem

\bibitem[\protect\citeauthoryear{{Fernandes}
  \textit{et~al.}}{2017}]{Fernandesetal:2017}
\begin{barticle}
\bauthor{\bsnm{{Fernandes}}, \binits{L.O.T.}},
\bauthor{\bsnm{{Kaufmann}}, \binits{P.}},
\bauthor{\bsnm{{Correia}}, \binits{E.}},
\bauthor{\bsnm{{Gim{\'e}nez de Castro}}, \binits{C.G.}},
\bauthor{\bsnm{{Kudaka}}, \binits{A.S.}},
\bauthor{\bsnm{{Marun}}, \binits{A.}},
\bauthor{\bsnm{{Pereyra}}, \binits{P.}},
\bauthor{\bsnm{{Raulin}}, \binits{J.-P.}},
\bauthor{\bsnm{{Valio}}, \binits{A.B.M.}}:
\byear{2017},
\batitle{{Spectral Trends of Solar Bursts at Sub-THz Frequencies}}.
\bjtitle{\solphys}
\bvolume{292},
\bfpage{21}.
\doiurl{10.1007/s11207-016-1043-6}.
\adsurl{2017SoPh..292...21F}.
\end{barticle}
\endbibitem

\bibitem[\protect\citeauthoryear{{Gim{\'e}nez de Castro}
  \textit{et~al.}}{2018}]{GimenezdeCastroetal:2018}
\begin{barticle}
\bauthor{\bsnm{{Gim{\'e}nez de Castro}}, \binits{C.G.}},
\bauthor{\bsnm{{Raulin}}, \binits{J.-P.}},
\bauthor{\bsnm{{Valle Silva}}, \binits{J.F.}},
\bauthor{\bsnm{{Sim{\~o}es}}, \binits{P.J.A.}},
\bauthor{\bsnm{{Kudaka}}, \binits{A.S.}},
\bauthor{\bsnm{{Valio}}, \binits{A.}}:
\byear{2018},
\batitle{{The 6 September 2017 X9 Super Flare Observed From Submillimeter to
  Mid-IR}}.
\bjtitle{Space Weather}
\bvolume{16}(\bissue{9}),
\bfpage{1261}.
\doiurl{10.1029/2018SW001969}.
\adsurl{https://ui.adsabs.harvard.edu/abs/2018SpWea..16.1261G}.
\end{barticle}
\endbibitem

\bibitem[\protect\citeauthoryear{{Heinzel} and
  {Avrett}}{2012}]{HeinzelAvrett:2012}
\begin{barticle}
\bauthor{\bsnm{{Heinzel}}, \binits{P.}},
\bauthor{\bsnm{{Avrett}}, \binits{E.H.}}:
\byear{2012},
\batitle{{Optical-to-Radio Continua in Solar Flares}}.
\bjtitle{\solphys}
\bvolume{277},
\bfpage{31}.
\doiurl{10.1007/s11207-011-9823-5}.
\adsurl{2012SoPh..277...31H}.
\end{barticle}
\endbibitem

\bibitem[\protect\citeauthoryear{{Kaufmann}
  \textit{et~al.}}{2004}]{Kaufmannetal:2004}
\begin{barticle}
\bauthor{\bsnm{{Kaufmann}}, \binits{P.}},
\bauthor{\bsnm{{Raulin}}, \binits{J.-P.}},
\bauthor{\bsnm{{de Castro}}, \binits{C.G.G.}},
\bauthor{\bsnm{{Levato}}, \binits{H.}},
\bauthor{\bsnm{{Gary}}, \binits{D.E.}},
\bauthor{\bsnm{{Costa}}, \binits{J.E.R.}},
\bauthor{\bsnm{{Marun}}, \binits{A.}},
\bauthor{\bsnm{{Pereyra}}, \binits{P.}},
\bauthor{\bsnm{{Silva}}, \binits{A.V.R.}},
\bauthor{\bsnm{{Correia}}, \binits{E.}}:
\byear{2004},
\batitle{{A New Solar Burst Spectral Component Emitting Only in the Terahertz
  Range}}.
\bjtitle{\apjl}
\bvolume{603}(\bissue{2}),
\bfpage{L121}.
\doiurl{10.1086/383186}.
\adsurl{https://ui.adsabs.harvard.edu/abs/2004ApJ...603L.121K}.
\end{barticle}
\endbibitem

\bibitem[\protect\citeauthoryear{{Kaufmann}
  \textit{et~al.}}{2008}]{Kaufmannetal:2008}
\begin{bchapter}
\bauthor{\bsnm{{Kaufmann}}, \binits{P.}},
\bauthor{\bsnm{{Levato}}, \binits{H.}},
\bauthor{\bsnm{{Cassiano}}, \binits{M.M.}},
\bauthor{\bsnm{{Correia}}, \binits{E.}},
\bauthor{\bsnm{{Costa}}, \binits{J.E.R.}},
\bauthor{\bsnm{{Gim{\'e}nez de Castro}}, \binits{C.G.}},
\bauthor{\bsnm{{Godoy}}, \binits{R.}},
\bauthor{\bsnm{{Kingsley}}, \binits{R.K.}},
\bauthor{\bsnm{{Kingsley}}, \binits{J.S.}},
\bauthor{\bsnm{{Kudaka}}, \binits{A.S.}},
\bauthor{\bsnm{{Marcon}}, \binits{R.}},
\bauthor{\bsnm{{Martin}}, \binits{R.}},
\bauthor{\bsnm{{Marun}}, \binits{A.}},
\bauthor{\bsnm{{Melo}}, \binits{A.M.}},
\bauthor{\bsnm{{Pereyra}}, \binits{P.}},
\bauthor{\bsnm{{Raulin}}, \binits{J.-P.}},
\bauthor{\bsnm{{Rose}}, \binits{T.}},
\bauthor{\bsnm{{Silva Valio}}, \binits{A.}},
\bauthor{\bsnm{{Walber}}, \binits{A.}},
\bauthor{\bsnm{{Wallace}}, \binits{P.}},
\bauthor{\bsnm{{Yakubovich}}, \binits{A.}},
\bauthor{\bsnm{{Zakia}}, \binits{M.B.}}:
\byear{2008},
\bctitle{{New telescopes for ground-based solar observations at submillimeter
  and mid-infrared}}.
In: \bbtitle{Society of Photo-Optical Instrumentation Engineers (SPIE)
  Conference Series}
\bseriesno{7012}.
\doiurl{10.1117/12.788889}.
\adsurl{2008SPIE.7012E..19K}.
\end{bchapter}
\endbibitem

\bibitem[\protect\citeauthoryear{{Kaufmann}
  \textit{et~al.}}{2009}]{Kaufmannetal:2009}
\begin{barticle}
\bauthor{\bsnm{{Kaufmann}}, \binits{P.}},
\bauthor{\bsnm{{Trottet}}, \binits{G.}},
\bauthor{\bsnm{{Gim{\'e}nez de Castro}}, \binits{C.G.}},
\bauthor{\bsnm{{Raulin}}, \binits{J.-P.}},
\bauthor{\bsnm{{Krucker}}, \binits{S.}},
\bauthor{\bsnm{{Shih}}, \binits{A.Y.}},
\bauthor{\bsnm{{Levato}}, \binits{H.}}:
\byear{2009},
\batitle{{Sub-terahertz, Microwaves and High Energy Emissions During the 6
  December 2006 Flare, at 18:40 UT}}.
\bjtitle{\solphys}
\bvolume{255}(\bissue{1}),
\bfpage{131}.
\doiurl{10.1007/s11207-008-9312-7}.
\adsurl{https://ui.adsabs.harvard.edu/abs/2009SoPh..255..131K}.
\end{barticle}
\endbibitem

\bibitem[\protect\citeauthoryear{{Kaufmann}
  \textit{et~al.}}{2013}]{Kaufmannetal:2013}
\begin{barticle}
\bauthor{\bsnm{{Kaufmann}}, \binits{P.}},
\bauthor{\bsnm{{White}}, \binits{S.M.}},
\bauthor{\bsnm{{Freeland}}, \binits{S.L.}},
\bauthor{\bsnm{{Marcon}}, \binits{R.}},
\bauthor{\bsnm{{Fernandes}}, \binits{L.O.T.}},
\bauthor{\bsnm{{Kudaka}}, \binits{A.S.}},
\bauthor{\bsnm{{de Souza}}, \binits{R.V.}},
\bauthor{\bsnm{{Aballay}}, \binits{J.L.}},
\bauthor{\bsnm{{Fernandez}}, \binits{G.}},
\bauthor{\bsnm{{Godoy}}, \binits{R.}},
\bauthor{\bsnm{{Marun}}, \binits{A.}},
\bauthor{\bsnm{{Valio}}, \binits{A.}},
\bauthor{\bsnm{{Raulin}}, \binits{J.-P.}},
\bauthor{\bsnm{{Gim{\'e}nez de Castro}}, \binits{C.G.}}:
\byear{2013},
\batitle{{A Bright Impulsive Solar Burst Detected at 30 THz}}.
\bjtitle{\apj}
\bvolume{768},
\bfpage{134}.
\doiurl{10.1088/0004-637X/768/2/134}.
\adsurl{2013ApJ...768..134K}.
\end{barticle}
\endbibitem

\bibitem[\protect\citeauthoryear{{Kaufmann}
  \textit{et~al.}}{2014}]{Kaufmannetal:2014}
\begin{barticle}
\bauthor{\bsnm{{Kaufmann}}, \binits{P.}},
\bauthor{\bsnm{{Marcon}}, \binits{R.}},
\bauthor{\bsnm{{Abrantes}}, \binits{A.}},
\bauthor{\bsnm{{Bortolucci}}, \binits{E.C.}},
\bauthor{\bsnm{{T. Fernandes}}, \binits{L.O.}},
\bauthor{\bsnm{{Kropotov}}, \binits{G.I.}},
\bauthor{\bsnm{{Kudaka}}, \binits{A.S.}},
\bauthor{\bsnm{{Machado}}, \binits{N.}},
\bauthor{\bsnm{{Marun}}, \binits{A.}},
\bauthor{\bsnm{{Nikolaev}}, \binits{V.}},
\bauthor{\bsnm{{Silva}}, \binits{A.}},
\bauthor{\bsnm{{da Silva}}, \binits{C.S.}},
\bauthor{\bsnm{{Timofeevsky}}, \binits{A.}}:
\byear{2014},
\batitle{{THz photometers for solar flare observations from space}}.
\bjtitle{Exp. Astron.}
\bvolume{37}(\bissue{3}),
\bfpage{579}.
\doiurl{10.1007/s10686-014-9389-y}.
\adsurl{https://ui.adsabs.harvard.edu/abs/2014ExA....37..579K}.
\end{barticle}
\endbibitem

\bibitem[\protect\citeauthoryear{{Kaufmann}
  \textit{et~al.}}{2015a}]{Kaufmannetal:2015}
\begin{barticle}
\bauthor{\bsnm{{Kaufmann}}, \binits{P.}},
\bauthor{\bsnm{{White}}, \binits{S.M.}},
\bauthor{\bsnm{{Marcon}}, \binits{R.}},
\bauthor{\bsnm{{Kudaka}}, \binits{A.S.}},
\bauthor{\bsnm{{Cabezas}}, \binits{D.P.}},
\bauthor{\bsnm{{Cassiano}}, \binits{M.M.}},
\bauthor{\bsnm{{Francile}}, \binits{C.}},
\bauthor{\bsnm{{Fernandes}}, \binits{L.O.T.}},
\bauthor{\bsnm{{Hidalgo Ramirez}}, \binits{R.F.}},
\bauthor{\bsnm{{Luoni}}, \binits{M.}},
\bauthor{\bsnm{{Marun}}, \binits{A.}},
\bauthor{\bsnm{{Pereyra}}, \binits{P.}},
\bauthor{\bsnm{{Souza}}, \binits{R.V.}}:
\byear{2015}a,
\batitle{{Bright 30 THz impulsive solar bursts}}.
\bjtitle{Journal of Geophysical Research (Space Physics)}
\bvolume{120},
\bfpage{4155}.
\doiurl{10.1002/2015JA021313}.
\adsurl{https://ui.adsabs.harvard.edu/\#abs/2015JGRA..120.4155K}.
\end{barticle}
\endbibitem

\bibitem[\protect\citeauthoryear{{Kaufmann}
  \textit{et~al.}}{2015b}]{Kaufmannetal:2015b}
\begin{bchapter}
\bauthor{\bsnm{{Kaufmann}}, \binits{P.}},
\bauthor{\bsnm{{Abrantes}}, \binits{A.}},
\bauthor{\bsnm{{Bortolucci}}, \binits{E.C.}},
\bauthor{\bsnm{{Fernandes}}, \binits{L.O.T.}},
\bauthor{\bsnm{I.}, \binits{K.G.}},
\bauthor{\bsnm{{Kudaka}}, \binits{A.S.}},
\bauthor{\bsnm{{Machado}}, \binits{N.}},
\bauthor{\bsnm{{Marcon}}, \binits{R.}},
\bauthor{\bsnm{{Nicolaev}}, \binits{V.}},
\bauthor{\bsnm{{Timojeevsky}}, \binits{A.}}:
\byear{2015}b,
\bctitle{Space and ground-based new tools for thz solar flare observations}.
In: \bbtitle{Proceedings of the 26$^{th}$ International Symposium on Space
  Terahertz Technology, ISSTT 2015}.
\bcomment{Link:
  \url{https://www.cfa.harvard.edu/events/2015/isstt2015/proceedings/}}.
\end{bchapter}
\endbibitem

\bibitem[\protect\citeauthoryear{{Ka{\v s}parov{\'a}}
  \textit{et~al.}}{2009}]{Kasparovaetal:2009}
\begin{barticle}
\bauthor{\bsnm{{Ka{\v s}parov{\'a}}}, \binits{J.}},
\bauthor{\bsnm{{Heinzel}}, \binits{P.}},
\bauthor{\bsnm{{Karlick{\'y}}}, \binits{M.}},
\bauthor{\bsnm{{Moravec}}, \binits{Z.}},
\bauthor{\bsnm{{Varady}}, \binits{M.}}:
\byear{2009},
\batitle{{Far-IR and Radio Thermal Continua in Solar Flares}}.
\bjtitle{Cent. Europ. Astrophys. Bull.}
\bvolume{33},
\bfpage{309}.
\adsurl{2009CEAB...33..309K}.
\end{barticle}
\endbibitem

\bibitem[\protect\citeauthoryear{{Krucker}
  \textit{et~al.}}{2013}]{Kruckeretal:2013}
\begin{barticle}
\bauthor{\bsnm{{Krucker}}, \binits{S.}},
\bauthor{\bsnm{{Gim{\'e}nez de Castro}}, \binits{C.G.}},
\bauthor{\bsnm{{Hudson}}, \binits{H.S.}},
\bauthor{\bsnm{{Trottet}}, \binits{G.}},
\bauthor{\bsnm{{Bastian}}, \binits{T.S.}},
\bauthor{\bsnm{{Hales}}, \binits{A.S.}},
\bauthor{\bsnm{{Ka{\v s}parov{\'a}}}, \binits{J.}},
\bauthor{\bsnm{{Klein}}, \binits{K.-L.}},
\bauthor{\bsnm{{Kretzschmar}}, \binits{M.}},
\bauthor{\bsnm{{L{\"u}thi}}, \binits{T.}},
\bauthor{\bsnm{{Mackinnon}}, \binits{A.}},
\bauthor{\bsnm{{Pohjolainen}}, \binits{S.}},
\bauthor{\bsnm{{White}}, \binits{S.M.}}:
\byear{2013},
\batitle{{Solar flares at submillimeter wavelengths}}.
\bjtitle{\aapr}
\bvolume{21},
\bfpage{58}.
\doiurl{10.1007/s00159-013-0058-3}.
\adsurl{2013A\%26ARv..21...58K}.
\end{barticle}
\endbibitem

\bibitem[\protect\citeauthoryear{{L{\"u}thi}, {Magun}, and
  {Miller}}{2004}]{Luthietal:2004a}
\begin{barticle}
\bauthor{\bsnm{{L{\"u}thi}}, \binits{T.}},
\bauthor{\bsnm{{Magun}}, \binits{A.}},
\bauthor{\bsnm{{Miller}}, \binits{M.}}:
\byear{2004},
\batitle{{First observation of a solar X-class flare in the submillimeter range
  with KOSMA}}.
\bjtitle{\aap}
\bvolume{415},
\bfpage{1123}.
\doiurl{10.1051/0004-6361:20034624}.
\adsurl{https://ui.adsabs.harvard.edu/abs/2004A&A...415.1123L}.
\end{barticle}
\endbibitem

\bibitem[\protect\citeauthoryear{{MacGregor}, {Osten}, and
  {Hughes}}{2020}]{MacGregoretal:2020}
\begin{barticle}
\bauthor{\bsnm{{MacGregor}}, \binits{A.M.}},
\bauthor{\bsnm{{Osten}}, \binits{R.A.}},
\bauthor{\bsnm{{Hughes}}, \binits{A.M.}}:
\byear{2020},
\batitle{{Properties of M Dwarf Flares at Millimeter Wavelengths}}.
\bjtitle{\apj}
\bvolume{891}(\bissue{1}),
\bfpage{80}.
\doiurl{10.3847/1538-4357/ab711d}.
\adsurl{https://ui.adsabs.harvard.edu/abs/2020ApJ...891...80M}.
\end{barticle}
\endbibitem

\bibitem[\protect\citeauthoryear{{MacGregor}
  \textit{et~al.}}{2018}]{MacGregoretal:2018}
\begin{barticle}
\bauthor{\bsnm{{MacGregor}}, \binits{M.A.}},
\bauthor{\bsnm{{Weinberger}}, \binits{A.J.}},
\bauthor{\bsnm{{Wilner}}, \binits{D.J.}},
\bauthor{\bsnm{{Kowalski}}, \binits{A.F.}},
\bauthor{\bsnm{{Cranmer}}, \binits{S.R.}}:
\byear{2018},
\batitle{{Detection of a Millimeter Flare from Proxima Centauri}}.
\bjtitle{\apjl}
\bvolume{855}(\bissue{1}),
\bfpage{L2}.
\doiurl{10.3847/2041-8213/aaad6b}.
\adsurl{https://ui.adsabs.harvard.edu/abs/2018ApJ...855L...2M}.
\end{barticle}
\endbibitem

\bibitem[\protect\citeauthoryear{{Machado}
  \textit{et~al.}}{1980}]{Machadoetal:1980}
\begin{barticle}
\bauthor{\bsnm{{Machado}}, \binits{M.E.}},
\bauthor{\bsnm{{Avrett}}, \binits{E.H.}},
\bauthor{\bsnm{{Vernazza}}, \binits{J.E.}},
\bauthor{\bsnm{{Noyes}}, \binits{R.W.}}:
\byear{1980},
\batitle{{Semiempirical models of chromospheric flare regions}}.
\bjtitle{\apj}
\bvolume{242},
\bfpage{336}.
\doiurl{10.1086/158467}.
\adsurl{1980ApJ...242..336M}.
\end{barticle}
\endbibitem

\bibitem[\protect\citeauthoryear{{Mauas}, {Machado}, and
  {Avrett}}{1990}]{Mauasetal:1990}
\begin{barticle}
\bauthor{\bsnm{{Mauas}}, \binits{P.J.D.}},
\bauthor{\bsnm{{Machado}}, \binits{M.E.}},
\bauthor{\bsnm{{Avrett}}, \binits{E.H.}}:
\byear{1990},
\batitle{{The white-light flare of 1982 June 15 - Models}}.
\bjtitle{\apj}
\bvolume{360},
\bfpage{715}.
\doiurl{10.1086/169157}.
\adsurl{1990ApJ...360..715M}.
\end{barticle}
\endbibitem

\bibitem[\protect\citeauthoryear{{Melo} \textit{et~al.}}{2008}]{Meloetal:2008}
\begin{barticle}
\bauthor{\bsnm{{Melo}}, \binits{A.M.}},
\bauthor{\bsnm{{Kornberg}}, \binits{M.A.}},
\bauthor{\bsnm{{Kaufmann}}, \binits{P.}},
\bauthor{\bsnm{{Piazzetta}}, \binits{M.H.}},
\bauthor{\bsnm{{Bortolucci}}, \binits{E.C.}},
\bauthor{\bsnm{{Zakia}}, \binits{M.B.}},
\bauthor{\bsnm{{Bauer}}, \binits{O.H.}},
\bauthor{\bsnm{{Poglitsch}}, \binits{A.}},
\bauthor{\bsnm{{Alves da Silva}}, \binits{A.M.P.}}:
\byear{2008},
\batitle{{Metal mesh resonant filters for terahertz frequencies}}.
\bjtitle{\appop}
\bvolume{47}(\bissue{32}),
\bfpage{6064}.
\doiurl{10.1364/AO.47.006064}.
\adsurl{https://ui.adsabs.harvard.edu/abs/2008ApOpt..47.6064M}.
\end{barticle}
\endbibitem

\bibitem[\protect\citeauthoryear{{Ohki} and {Hudson}}{1975}]{OhkiHudson:1975}
\begin{barticle}
\bauthor{\bsnm{{Ohki}}, \binits{K.}},
\bauthor{\bsnm{{Hudson}}, \binits{H.S.}}:
\byear{1975},
\batitle{{The solar-flare infrared continuum}}.
\bjtitle{\solphys}
\bvolume{43},
\bfpage{405}.
\adsurl{cgi-bin/nph-bib_query?bibcode=1975SoPh...43..405O&amp;db_key=AST}.
\end{barticle}
\endbibitem

\bibitem[\protect\citeauthoryear{{Pardo}, {Cernicharo}, and
  {Serabyn}}{2001}]{Pardoetal:2001}
\begin{barticle}
\bauthor{\bsnm{{Pardo}}, \binits{J.R.}},
\bauthor{\bsnm{{Cernicharo}}, \binits{J.}},
\bauthor{\bsnm{{Serabyn}}, \binits{E.}}:
\byear{2001},
\batitle{{Atmospheric transmission at microwaves (ATM): an improved model for
  millimeter/submillimeter applications}}.
\bjtitle{IEEE Trans. on Antenn. Propag.}
\bvolume{49}(\bissue{12}),
\bfpage{1683}.
\doiurl{10.1109/8.982447}.
\adsurl{https://ui.adsabs.harvard.edu/abs/2001ITAP...49.1683P}.
\end{barticle}
\endbibitem

\bibitem[\protect\citeauthoryear{{Penn} \textit{et~al.}}{2016}]{Pennetal:2016}
\begin{barticle}
\bauthor{\bsnm{{Penn}}, \binits{M.}},
\bauthor{\bsnm{{Krucker}}, \binits{S.}},
\bauthor{\bsnm{{Hudson}}, \binits{H.}},
\bauthor{\bsnm{{Jhabvala}}, \binits{M.}},
\bauthor{\bsnm{{Jennings}}, \binits{D.}},
\bauthor{\bsnm{{Lunsford}}, \binits{A.}},
\bauthor{\bsnm{{Kaufmann}}, \binits{P.}}:
\byear{2016},
\batitle{{Spectral and Imaging Observations of a White-light Solar Flare in the
  Mid-infrared}}.
\bjtitle{\apjl}
\bvolume{819},
\bfpage{L30}.
\doiurl{10.3847/2041-8205/819/2/L30}.
\adsurl{2016ApJ...819L..30P}.
\end{barticle}
\endbibitem

\bibitem[\protect\citeauthoryear{{Ramaty}}{1969}]{Ramaty:1969}
\begin{barticle}
\bauthor{\bsnm{{Ramaty}}, \binits{R.}}:
\byear{1969},
\batitle{{Gyrosynchrotron Emission and Absorption in a Magnetoactive Plasma}}.
\bjtitle{\apj}
\bvolume{158},
\bfpage{753}.
\doiurl{10.1086/150235}.
\adsurl{https://ui.adsabs.harvard.edu/abs/1969ApJ...158..753R}.
\end{barticle}
\endbibitem

\bibitem[\protect\citeauthoryear{{Righini} and
  {Simon}}{1976}]{RighiniSimon:1976}
\begin{barticle}
\bauthor{\bsnm{{Righini}}, \binits{G.}},
\bauthor{\bsnm{{Simon}}, \binits{M.}}:
\byear{1976},
\batitle{{Solar brightness temperature distribution at 350 and 450 microns.}}
\bjtitle{\apjl}
\bvolume{203},
\bfpage{L95}.
\doiurl{10.1086/182027}.
\adsurl{https://ui.adsabs.harvard.edu/abs/1976ApJ...203L..95R}.
\end{barticle}
\endbibitem

\bibitem[\protect\citeauthoryear{{Silva}
  \textit{et~al.}}{2007}]{Silvaetal:2007}
\begin{barticle}
\bauthor{\bsnm{{Silva}}, \binits{A.V.R.}},
\bauthor{\bsnm{{Share}}, \binits{G.H.}},
\bauthor{\bsnm{{Murphy}}, \binits{R.J.}},
\bauthor{\bsnm{{Costa}}, \binits{J.E.R.}},
\bauthor{\bsnm{{de Castro}}, \binits{C.G.G.}},
\bauthor{\bsnm{{Raulin}}, \binits{J.-P.}},
\bauthor{\bsnm{{Kaufmann}}, \binits{P.}}:
\byear{2007},
\batitle{{Evidence that Synchrotron Emission from Nonthermal Electrons Produces
  the Increasing Submillimeter Spectral Component in Solar Flares}}.
\bjtitle{\solphys}
\bvolume{245},
\bfpage{311}.
\doiurl{10.1007/s11207-007-9044-0}.
\adsurl{2007SoPh..245..311S}.
\end{barticle}
\endbibitem

\bibitem[\protect\citeauthoryear{{Sim{\~o}es}
  \textit{et~al.}}{2017}]{Simoesetal:2017}
\begin{barticle}
\bauthor{\bsnm{{Sim{\~o}es}}, \binits{P.J.A.}},
\bauthor{\bsnm{{Kerr}}, \binits{G.S.}},
\bauthor{\bsnm{{Fletcher}}, \binits{L.}},
\bauthor{\bsnm{{Hudson}}, \binits{H.S.}},
\bauthor{\bsnm{{Gim{\'e}nez de Castro}}, \binits{C.G.}},
\bauthor{\bsnm{{Penn}}, \binits{M.}}:
\byear{2017},
\batitle{{Formation of the thermal infrared continuum in solar flares}}.
\bjtitle{\aap}
\bvolume{605},
\bfpage{A125}.
\doiurl{10.1051/0004-6361/201730856}.
\adsurl{https://ui.adsabs.harvard.edu/abs/2017A&A...605A.125S}.
\end{barticle}
\endbibitem

\bibitem[\protect\citeauthoryear{{Trottet}
  \textit{et~al.}}{2015}]{Trottetetal:2015}
\begin{barticle}
\bauthor{\bsnm{{Trottet}}, \binits{G.}},
\bauthor{\bsnm{{Raulin}}, \binits{J.-P.}},
\bauthor{\bsnm{{Mackinnon}}, \binits{A.}},
\bauthor{\bsnm{{Gim{\'e}nez de Castro}}, \binits{G.}},
\bauthor{\bsnm{{Sim{\~o}es}}, \binits{P.J.A.}},
\bauthor{\bsnm{{Cabezas}}, \binits{D.}},
\bauthor{\bsnm{{de La Luz}}, \binits{V.}},
\bauthor{\bsnm{{Luoni}}, \binits{M.}},
\bauthor{\bsnm{{Kaufmann}}, \binits{P.}}:
\byear{2015},
\batitle{{Origin of the 30 THz Emission Detected During the Solar Flare on 2012
  March 13 at 17:20 UT}}.
\bjtitle{\solphys}
\bvolume{290},
\bfpage{2809}.
\doiurl{{10.1007/s11207-015-0782-0}}.
\adsurl{2015SoPh..290.2809T}.
\end{barticle}
\endbibitem

\bibitem[\protect\citeauthoryear{{Valio}
  \textit{et~al.}}{2013}]{Valioetal:2013}
\begin{barticle}
\bauthor{\bsnm{{Valio}}, \binits{A.}},
\bauthor{\bsnm{{Kaufmann}}, \binits{P.}},
\bauthor{\bsnm{{Gim{\'e}nez de Castro}}, \binits{C.G.}},
\bauthor{\bsnm{{Raulin}}, \binits{J.-P.}},
\bauthor{\bsnm{{Fernandes}}, \binits{L.O.T.}},
\bauthor{\bsnm{{Marun}}, \binits{A.}}:
\byear{2013},
\batitle{{POlarization Emission of Millimeter Activity at the Sun (POEMAS): New
  Circular Polarization Solar Telescopes at Two Millimeter Wavelength Ranges}}.
\bjtitle{\solphys}
\bvolume{283},
\bfpage{651}.
\doiurl{10.1007/s11207-013-0237-4}.
\adsurl{2013SoPh..283..651V}.
\end{barticle}
\endbibitem

\end{thebibliography}

\end{article}
\end{document}